\documentclass[11pt,a4paper]{article}
\usepackage{fullpage,graphicx,xcolor}

\newtheorem{theorem}{Theorem}
\newtheorem{lemma}[theorem]{Lemma}

\newcommand{\BWT}{\ensuremath{\mathrm{BWT}}}

\begin{document}

\title{Patching leaky tunnels in BWTs}
\author{Paola Bonizzoni, Davide Cozzi, Travis Gagie,\\
Younan Gao and Ragnar Groot Koerkamp}
\maketitle

\begin{abstract}
\noindent
We extend Baier's foundational work on tunnelling Burrows-Wheeler Transforms (BWTs) by showing how something that would be a good tunnel except for a few strings that diverge from the tunnel (``leaks''), can be patched and turned into a tunnel.  Whereas augmenting a BWT with a normal tunnel takes extra space proportional to the tunnels length, augmenting it with a patched tunnel takes that extra space plus space proportional to the number of leaks.
\end{abstract}

\section{Introduction}
\label{sec:introduction}

Suppose a string terminated by a special end-of-string symbol {\tt \$} lexicographically less than any other symbol.  The Burrows-Wheeler Transform (BWT)~\cite{BW94} of the string is obtained by sorting its characters into the lexicographic order of the suffixes that follow them, considering the string as circular (so the suffix following the {\tt \$} is the string itself).  For example, as shown on the left in Figure~\ref{fig:small_example}, the BWT of {\tt GATTACAT\$} is {\tt TTCGA\$ATA}: if we sort the cyclic shifts lexicographically, the BWT is the concatenation of the final character of each shift (so the contents of the last column in the matrix).  The extended BWT (eBWT)~\cite{MRRS07} of several such strings is obtained by sorting all the strings' characters into the lexicographic order of the cyclic shifts starting at the next characters, repeated infinitely.  For example, as shown on the right in Figure~\ref{fig:small_example}, the eBWT of {\tt GATTACAT\$}, {\tt GATACAT\$} and {\tt GATTAGATA\$} is {\tt 
TTATTTTCCGGGGAAA\$\$\$AAAATTAA}.  The BWTs of a repetitive string or the eBWTs of many similar strings are usually run-length compressible and researchers working with such strings often use run-length compressed BWTs (RLBWTs), extended or not.  For the sake of simplicity, henceforth we refer to both BWTs and eBWTs simply as BWTs.

\begin{figure}[t]
\begin{center}
{\tt
\begin{tabular}{c@{\hspace{10ex}}c}
\begin{tabular}{l}
\$GATTACAT \\
ACAT\$GATT \\
AT\$GATTAC \\
ATTACAT\$G \\
CAT\$GATTA \\
GATTACAT\$ \\
T\$GATTACA \\
TACAT\$GAT \\
TTACAT\$GA
\end{tabular}
&
\begin{tabular}{cl}
T  & \$GATACAT\$GATACAT... \\
T  & \$GATTACAT\$GATTACAT... \\
A  & \$GATTAGATA\$GATTAGATA... \\
T  & A\$GATTAGATA\$GATTAGAT... \\
T  & ACAT\$GATACAT\$GAT... \\
T  & ACAT\$GATTACAT\$GATT... \\
T  & AGATA\$GATTAGATA\$GATT... \\
C  & AT\$GATACAT\$GATAC... \\
C  & AT\$GATTACAT\$GATTAC... \\
G  & ATA\$GATTAGATA\$GATTAG... \\
G  & ATACAT\$GATACAT\$G... \\
G  & ATTACAT\$GATTACAT\$G... \\
G  & ATTAGATA\$GATTAGATA\$G... \\
A  & CAT\$GATACAT\$GATA... \\
A  & CAT\$GATTACAT\$GATTA... \\
A  & GATA\$GATTAGATA\$GATTA... \\
\$ & GATACAT\$GATACAT\$... \\
\$ & GATTACAT\$GATTACAT\$... \\
\$ & GATTAGATA\$GATTAGATA\$ \\
A  & T\$GATACAT\$GATACA... \\
A  & T\$GATTACAT\$GATTACA... \\
A  & TA\$GATTAGATA\$GATTAGA... \\
A  & TACAT\$GATACAT\$GA... \\
T  & TACAT\$GATTACAT\$GAT... \\
T  & TAGATA\$GATTAGATA\$GAT... \\
A  & TTACAT\$GATTACAT\$GA... \\
A  & TTAGATA\$GATTAGATA\$GA...
\end{tabular}
\end{tabular}}
\caption{On the left, the cyclic shifts of {\tt GATTACAT\$} in lexicographic order, with its BWT {\tt TTCGA\$ATA} in the last column of the matrix.  On the right, the eBWT {\tt 
TTATTTTCCGGGGAAA\$\$\$AAAATTAA} of {\tt GATTACAT\$}, {\tt GATACAT\$} and {\tt GATTAGATA\$}: the characters in those strings sorted in the lexicographic order of the cyclic shifts starting at the next characters, repeated infinitely.}
\label{fig:small_example}
\end{center}
\end{figure}

Researchers are interested in BWTs mainly because Ferragina and Manzini~\cite{FM05} showed how to use them for indexing.  For example, suppose we want to count the occurrences of {\tt TTA} in {\tt GATTACAT\$}, {\tt GATACAT\$} and {\tt GATTAGATA\$}.  We can easily find the interval of the BWT of those strings containing characters immediately preceding occurrences of {\tt A} in their string, because the starting position of that interval (counting from 1) is 1 plus the sum of the frequencies of characters lexicographically smaller than {\tt A} and its ending position is the sum of those frequencies plus the frequency of {\tt A} itself.  That interval --- the 4th through 13th lines of the right-hand side of Figure~\ref{fig:small_example} --- contains the 3rd through 6th copies of {\tt T} in the BWT.  It follows that the starting position of the interval of the BWT containing characters immediately preceding occurrences of {\tt TA}, is the sum of the frequencies of characters lexicographically smaller than {\tt T} plus 3, and its ending position is the sum of the those frequencies plus 6.  That interval --- the 22nd through 25th lines --- contains the 7th and 8th copies of {\tt T} in the BWT.  Therefore, there are 2 occurrences of {\tt TTA} in {\tt GATTACAT\$}, {\tt GATACAT\$} and {\tt GATTAGATA\$}.  This process of finding BWT intervals by processing a pattern from right to left is called backward stepping.

The most efficient indexes based on RLBWTs over small alphabets now use table lookup for backward stepping (see, e.g.,~\cite{ZBGL26}).  These demonstrate better locality than older implementations that used rank queries, but they still perform a non-trivial computation to find each BWT interval.  This is wasteful when each BWT intervals we find contains only copies of the character we are considering in the pattern.  For example, for this set of strings (with the {\tt -}'s shown only to illustrate alignment)
\begin{center}
{\tt \begin{tabular}{l}
AAGGACTCGG-ACGTTGCAAGTGGT\$ \\
ACCGACTCGG-ACGTTGCAAGTTCA\$ \\
ACCGACTCGG-A-GTTGCAATCTCC\$ \\
CGTGACTCGG-ACGTTGCAATCACT\$ \\
CGTGACTCGG-ACGTTGCAATGCCT\$ \\
CTTGACTCGG-ACGTTGCAATGTTT\$ \\
CTTGACTCGGGACGTTGCAATTGCT\$ \\
CTTGACTCGG-TCGTTGCAATTGCT\$
\end{tabular}}
\end{center}
the BWT is shown in Figure~\ref{fig:big_example}.  If we are looking for the BWT interval containing the characters preceding occurrences of {\tt GGACGTTGCAAT} and we have already found the interval $\BWT [40..45]$ containing the characters preceding occurrences of {\tt AT}, then all of the BWT intervals we find are unary until the one $\BWT [141..146]$ containing the characters preceding occurrences of {\tt GTTGCAAT}.

\begin{figure}[t]
\resizebox{\textwidth}{!}
{\tt \begin{tabular}{c@{\hspace{4ex}}c@{\hspace{4ex}}c@{\hspace{4ex}}c}
\begin{tabular}{rcl}
{\rm 1} & T & \$AA... \\
{\rm 2} & A & \$ACCGACTCGGAC... \\
{\rm 3} & C & \$ACCGACTCGGAG... \\
{\rm 4} & T & \$CGTGACTCGGACGTTGCAATC... \\
{\rm 5} & T & \$CGTGACTCGGACGTTGCAATG... \\
{\rm 6} & T & \$CTTGACTCGGA... \\
{\rm 7} & T & \$CTTGACTCGGG... \\
{\rm 8} & T & \$CTTGACTCGGT... \\
{\rm 9} & C & A\$... \\
{\rm 10} & \$ & AAGG... \\
{\rm 11} & C & AAGTG... \\
{\rm 12} & C & AAGTT... \\
{\rm 13} & C & AATCA... \\
{\rm 14} & C & AATCT... \\
{\rm 15} & C & AATGC... \\
{\rm 16} & C & AATGT... \\
{\rm 17} & C & AATTGCT\$CTTGACTCGGG... \\
{\rm 18} & C & AATTGCT\$CTTGACTCGGT... \\
{\rm 19} & \$ & ACCGACTCGGAC... \\
{\rm 20} & \$ & ACCGACTCGGAG... \\
{\rm 21} & G & ACGTTGCAAGTG... \\
{\rm 22} & G & ACGTTGCAAGTT... \\
{\rm 23} & G & ACGTTGCAATC... \\
{\rm 24} & G & ACGTTGCAATGC... \\
{\rm 25} & G & ACGTTGCAATGT... \\
{\rm 26} & G & ACGTTGCAATT... \\
{\rm 27} & C & ACT\$... \\
{\rm 28} & G & ACTCGGACGTTGCAAGTG... \\
{\rm 29} & G & ACTCGGACGTTGCAAGTT... \\
{\rm 30} & G & ACTCGGACGTTGCAATC... \\
{\rm 31} & G & ACTCGGACGTTGCAATGC... \\
{\rm 32} & G & ACTCGGACGTTGCAATGT... \\
{\rm 33} & G & ACTCGGAG... \\
{\rm 34} & G & ACTCGGG... \\
{\rm 35} & G & ACTCGGT... \\
{\rm 36} & A & AGG... \\
{\rm 37} & A & AGTG... \\
{\rm 38} & A & AGTTC... \\
{\rm 39} & G & AGTTG... \\
{\rm 40} & A & ATCA... \\
{\rm 41} & A & ATCT... \\
{\rm 42} & A & ATGC... \\
{\rm 43} & A & ATGT... \\
{\rm 44} & A & ATTGCT\$CTTGACTCGGG... \\
{\rm 45} & A & ATTGCT\$CTTGACTCGGT... \\
{\rm 46} & C & C\$... \\
{\rm 47} & T & CA\$... \\
{\rm 48} & G & CAAGTG... \\
{\rm 49} & G & CAAGTT... \\
{\rm 50} & G & CAATCA...
\end{tabular}
&
\begin{tabular}{rcl}
{\rm 51} & G & CAATCT... \\
{\rm 52} & G & CAATGC... \\
{\rm 53} & G & CAATGT... \\
{\rm 54} & G & CAATTGCT\$CTTGACTCGGG... \\
{\rm 55} & G & CAATTGCT\$CTTGACTCGGT... \\
{\rm 56} & T & CAC... \\
{\rm 57} & T & CC\$... \\
{\rm 58} & A & CCGACTCGGAC... \\
{\rm 59} & A & CCGACTCGGAG... \\
{\rm 60} & G & CCT... \\
{\rm 61} & C & CGACTCGGAC... \\
{\rm 62} & C & CGACTCGGAG... \\
{\rm 63} & T & CGGACGTTGCAAGTG... \\
{\rm 64} & T & CGGACGTTGCAAGTT... \\
{\rm 65} & T & CGGACGTTGCAATC... \\
{\rm 66} & T & CGGACGTTGCAATGC... \\
{\rm 67} & T & CGGACGTTGCAATGT... \\
{\rm 68} & T & CGGAG... \\
{\rm 69} & T & CGGG... \\
{\rm 70} & T & CGGT... \\
{\rm 71} & \$ & CGTGACTCGGACGTTGCAATC... \\
{\rm 72} & \$ & CGTGACTCGGACGTTGCAATG... \\
{\rm 73} & A & CGTTGCAAGTG... \\
{\rm 74} & A & CGTTGCAAGTT... \\
{\rm 75} & A & CGTTGCAATC... \\
{\rm 76} & A & CGTTGCAATGC... \\
{\rm 77} & A & CGTTGCAATGT... \\
{\rm 78} & A & CGTTGCAATTGCT\$CTTGACTCGGG... \\
{\rm 79} & T & CGTTGCAATTGCT\$CTTGACTCGGT... \\
{\rm 80} & A & CT\$CGTGACTCGGACGTTGCAATC... \\
{\rm 81} & C & CT\$CGTGACTCGGACGTTGCAATG... \\
{\rm 82} & G & CT\$CTTGACTCGGG... \\
{\rm 83} & G & CT\$CTTGACTCGGT... \\
{\rm 84} & T & CTCC... \\
{\rm 85} & A & CTCGGACGTTGCAAGTG... \\
{\rm 86} & A & CTCGGACGTTGCAAGTT... \\
{\rm 87} & A & CTCGGACGTTGCAATC... \\
{\rm 88} & A & CTCGGACGTTGCAATGC... \\
{\rm 89} & A & CTCGGACGTTGCAATGT... \\
{\rm 90} & A & CTCGGAG... \\
{\rm 91} & A & CTCGGG... \\
{\rm 92} & A & CTCGGT... \\
{\rm 93} & \$ & CTTGACTCGGA... \\
{\rm 94} & \$ & CTTGACTCGGG... \\
{\rm 95} & \$ & CTTGACTCGGT... \\
{\rm 96} & G & GACGTTGCAAGTG... \\
{\rm 97} & G & GACGTTGCAAGTT... \\
{\rm 98} & G & GACGTTGCAATC... \\
{\rm 99} & G & GACGTTGCAATGC... \\
{\rm 100} & G & GACGTTGCAATGT...
\end{tabular}
&
\begin{tabular}{rcl}
{\rm 101} & G & GACGTTGCAATT... \\
{\rm 102} & G & GACTCGGACGTTGCAAGTG... \\
{\rm 103} & C & GACTCGGACGTTGCAAGTT... \\
{\rm 104} & T & GACTCGGACGTTGCAATC... \\
{\rm 105} & T & GACTCGGACGTTGCAATGC... \\
{\rm 106} & T & GACTCGGACGTTGCAATGT... \\
{\rm 107} & C & GACTCGGAG... \\
{\rm 108} & T & GACTCGGG... \\
{\rm 109} & T & GACTCGGT... \\
{\rm 110} & G & GAG... \\
{\rm 111} & T & GCAAGTG... \\
{\rm 112} & T & GCAAGTT... \\
{\rm 113} & T & GCAATCA... \\
{\rm 114} & T & GCAATCT... \\
{\rm 115} & T & GCAATGC... \\
{\rm 116} & T & GCAATGT... \\
{\rm 117} & T & GCAATTGCT\$CTTGACTCGGG... \\
{\rm 118} & T & GCAATTGCT\$CTTGACTCGGT... \\
{\rm 119} & T & GCC... \\
{\rm 120} & T & GCT\$CTTGACTCGGG... \\
{\rm 121} & T & GCT\$CTTGACTCGGT... \\
{\rm 122} & C & GGACGTTGCAAGTG... \\
{\rm 123} & C & GGACGTTGCAAGTT... \\
{\rm 124} & C & GGACGTTGCAATC... \\
{\rm 125} & C & GGACGTTGCAATGC... \\
{\rm 126} & C & GGACGTTGCAATGT... \\
{\rm 127} & G & GGACGTTGCAATT... \\
{\rm 128} & A & GGACT... \\
{\rm 129} & C & GGAG... \\
{\rm 130} & C & GGG... \\
{\rm 131} & T & GGT\$... \\
{\rm 132} & C & GGTC... \\
{\rm 133} & G & GT\$... \\
{\rm 134} & G & GTC... \\
{\rm 135} & C & GTGACTCGGACGTTGCAATC... \\
{\rm 136} & C & GTGACTCGGACGTTGCAATG... \\
{\rm 137} & A & GTGG... \\
{\rm 138} & A & GTTC... \\
{\rm 139} & C & GTTGCAAGTG... \\
{\rm 140} & C & GTTGCAAGTT... \\
{\rm 141} & C & GTTGCAATCA... \\
{\rm 142} & A & GTTGCAATCT... \\
{\rm 143} & C & GTTGCAATGC... \\
{\rm 144} & C & GTTGCAATGT... \\
{\rm 145} & C & GTTGCAATTGCT\$CTTGACTCGGG... \\
{\rm 146} & C & GTTGCAATTGCT\$CTTGACTCGGT... \\
{\rm 147} & T & GTTT... \\
{\rm 148} & G & T\$A... \\
{\rm 149} & C & T\$CGTGACTCGGACGTTGCAATC... \\
{\rm 150} & C & T\$CGTGACTCGGACGTTGCAATG...
\end{tabular}
&
\begin{tabular}{rcl}
{\rm 151} & T & T\$CTTGACTCGGA... \\
{\rm 152} & C & T\$CTTGACTCGGG... \\
{\rm 153} & C & T\$CTTGACTCGGT... \\
{\rm 154} & T & TCA\$... \\
{\rm 155} & A & TCAC... \\
{\rm 156} & C & TCC... \\
{\rm 157} & C & TCGGACGTTGCAAGTG... \\
{\rm 158} & C & TCGGACGTTGCAAGTT... \\
{\rm 159} & C & TCGGACGTTGCAATC... \\
{\rm 160} & C & TCGGACGTTGCAATGC... \\
{\rm 161} & C & TCGGACGTTGCAATGT... \\
{\rm 162} & C & TCGGAG... \\
{\rm 163} & C & TCGGG... \\
{\rm 164} & C & TCGGT... \\
{\rm 165} & G & TCGT... \\
{\rm 166} & A & TCT... \\
{\rm 167} & G & TGACTCGGACGTTGCAATC... \\
{\rm 168} & G & TGACTCGGACGTTGCAATGC... \\
{\rm 169} & T & TGACTCGGACGTTGCAATGT... \\
{\rm 170} & T & TGACTCGGG... \\
{\rm 171} & T & TGACTCGGT... \\
{\rm 172} & T & TGCAAGTG... \\
{\rm 173} & T & TGCAAGTT... \\
{\rm 174} & T & TGCAATCA... \\
{\rm 175} & T & TGCAATCT... \\
{\rm 176} & T & TGCAATGC... \\
{\rm 177} & T & TGCAATGT... \\
{\rm 178} & T & TGCAATTGCT\$CTTGACTCGGG... \\
{\rm 179} & T & TGCAATTGCT\$CTTGACTCGGT... \\
{\rm 180} & A & TGCC... \\
{\rm 181} & T & TGCT\$CTTGACTCGGG... \\
{\rm 182} & T & TGCT\$CTTGACTCGGT... \\
{\rm 183} & G & TGG... \\
{\rm 184} & A & TGT... \\
{\rm 185} & T & TT\$... \\
{\rm 186} & G & TTC... \\
{\rm 187} & C & TTGACTCGGA... \\
{\rm 188} & C & TTGACTCGGG... \\
{\rm 189} & C & TTGACTCGGT... \\
{\rm 190} & G & TTGCAAGTG... \\
{\rm 191} & G & TTGCAAGTT... \\
{\rm 192} & G & TTGCAATCA... \\
{\rm 193} & G & TTGCAATCT... \\
{\rm 194} & G & TTGCAATGC... \\
{\rm 195} & G & TTGCAATGT... \\
{\rm 196} & G & TTGCAATTGCT\$CTTGACTCGGG... \\
{\rm 197} & G & TTGCAATTGCT\$CTTGACTCGGT... \\
{\rm 198} & A & TTGCT\$CTTGACTCGGG... \\
{\rm 199} & A & TTGCT\$CTTGACTCGGT... \\
{\rm 200} & G & TTT...
\end{tabular}
\end{tabular}}
\caption{The BWT for {\tt AAGGACTCGGACGTTGCAAGTGGT\$}, {\tt ACCGACTCGGACGTTGCAAGTTCA\$}, {\tt ACCGACTCGGAGTTGCAATCTCC\$}, {\tt CGTGACTCGGACGTTGCAATCACT\$}, {\tt CGTGACTCGGACGTTGCAATGCCT\$}, {\tt CTTGACTCGGACGTTGCAATGTTT\$}, {\tt CTTGACTCGGGACGTTGCAATTGCT\$}, {\tt CTTGACTCGGTCGTTGCAATTGCT\$}, with enough of each repeated cyclic shift shown to distinguish it from its lexicographic predecessor and successor.}
\label{fig:big_example}
\end{figure}

We can skip directly from the interval $\BWT [40..45]$ to the interval $\BWT [141..146]$ if we store the pair $(40, 45)$ in a hash table with satellite data {\tt GTTGCA} and 141 (the 146 is superfluous because the intervals are the same length).   We just need to compare the stored string to the corresponding substring of our pattern, which is much faster than backward stepping.  Baier~\cite{Bai18} was the first to study such shortcuts through the BWT, which he called tunnels (although we take some artistic license with his definition).  If we store as satellite data also 13, 50, 113, 174 and 192 --- encoding the BWT intervals containing the characters preceding occurrences of {\tt AAT}, {\tt CAAT}, {\tt GCAAT}, {\tt TGCAAT} and {\tt TTGCAAT} --- then, even when our pattern diverges from the tunnel somewhere in the middle, we can still use part of the tunnel as a shortcut.

If we are looking for the BWT interval containing the characters preceding occurrences of {\tt GGACGTTGCAATG} then after we have found the interval $\BWT [42..43]$ containing characters preceding occurrences of {\tt ATG}, we can use the tunnel to skip to the interval $\BWT [143..144]$ containing the characters preceding occurrences of {\tt GTTGCAATG}.  To do this, instead of storing $(40, 45)$ in a hash table, we store as a point on a grid supporting left-above dominance queries, so the query $(42, 43)$ finds it (with the same satellite data).  Once we have $(40, 45)$, we compute $143 = 141 + (42 - 40)$ and $144 = 141 + (43 - 40)$.  If our pattern diverges from the tunnel then we add the offset $42 - 40 = 2$ to the appropriate pointer (either 13, 50, 113, 174 or 192).  If we also store all the (implicit) points in the satellite data on the grid, then we can join a tunnel in the middle.  This could be useful when our pattern ends with a substring in the middle of the string associated with the tunnel --- called its label --- that is specific to the occurrences of that label.

\begin{lemma}
We can augment a BWT-based index with a tunnel at the cost of using extra space proportional to the tunnel's length.  Checking whether we can use a tunnel after a backward step takes $O \left( \frac{\log n}{\log \log n} \right)$ time.  Following a tunnel until we reach the end or our pattern diverges from it requires only time proportional to what is needed to find the length of the longest common suffix of the appropriate prefixes of our pattern and the tunnel's label (which depends on the format in which the pattern is given and the label stored, but is generally much less than the time to backward step through that longest common suffix).  After that, dropping back into the BWT takes constant time.
\end{lemma}

Baier and Dede~\cite{BD19} showed that tunnelling optimally is NP-hard when the tunnels can overlap but takes polynomial time when they must be disjoint, and there have been practical papers (such as~\cite{BSL26}) that use tunnelling.  Most of those papers are motivated by pangenomics, but tunnelling has a scaling problem when applied to datasets of thousands or millions of genomes: if even one individual has a variation in the middle of what would otherwise be a wonderful tunnel --- a leak, to put it colloquially --- it breaks the tunnel.  In our example, we cannot have a tunnel for a string ending {\tt GACT} because 5 of its occurrences are preceded by {\tt T}s but 2 of them by {\tt C}s and 1 by a {\tt G}.

With enough individuals in the dataset, there are so many rare variations littered throughout their genomes that good tunnels --- both wide and long --- without leaks become rare.  Since everyone carries rare variations we cannot solve this problem by just excluding some genomes from the dataset, and that would anyway defeat one of the main purposes of pangenomics: to better reflect genetic diversity.  In this paper we consider whether we can patch leaky tunnels and still use them, and show that it is actually very easy, at least in theory.  We leave as future work implementing and testing our scheme in practice.  The most promising application is letting us often skip over regions of many consecutive sparse columns in haplotype panels when we are searching with Durbin's~\cite{Dur14} positional BWT (PBWT).  We are also interested in generalizing our scheme to patching leaky tunnels in Wheeler Graphs~\cite{GMS17,AGNS19,GB22}.

\section{Patches}
\label{sec:patches}

Suppose we have found the interval $\BWT [48..55]$ containing characters preceding occurrences of {\tt CAA} in our example.  Five of its 8 occurrences are preceded by {\tt GACTCGGACGTTG}, so we might want a tunnel to the interval $\BWT [102..106]$ containing the characters preceding occurrences of {\tt GACTCGGACGTTGCAA}.  Unfortunately, there are 3 leaks in this tunnel: a missing {\tt C} in {\tt ACCGACTCGGAGTTGCAATCTCC\$}, an {\tt A}-to-{\tt T} substitution in {\tt CTTGACTCGGTCGTTGCAATTGCT\$}, and an extra {\tt G} in {\tt CTTGACTCGGGACGTTGCAATTGCT\$}.  We might be able to deal with the {\tt A}-to-{\tt T} substition and the extra {\tt G} just by narrowing the mouth of the tunnel, since those leaks are at the ends of their respective BWT intervals (the {\tt T} at $\BWT [79]$ at the end of the interval $\BWT [73..79]$ for {\tt CGTTGCAA} and the {\tt G} at $\BWT [127]$ at the end of the interval $\BWT [122..127]$ for {\tt GGACGTTGCAA}), but the leak for the missing {\tt C} is right in the middle of the corresponding BWT interval (the {\tt A} at $\BWT [142]$ in the interval $\BWT [139..146]$ for {\tt GTTGCAA}).

To patch all these leaks at once, we consider the occurrences of {\tt CAA} in the lexicographic order of the repeated cyclic shifts that follow them:
\begin{center}
{\tt \begin{tabular}{l}
\textcolor{white}{\ GACTCGGACGTTG}CAAGTG... \\
\textcolor{white}{\ GACTCGGACGTTG}CAAGTT... \\
\textcolor{white}{\ GACTCGGACGTTG}CAATCA... \\
\textcolor{white}{\ \ GACTCGGAGTTG}CAATCT... \\
\textcolor{white}{\ GACTCGGACGTTG}CAATGC... \\
\textcolor{white}{\ GACTCGGACGTTG}CAATGT... \\
\textcolor{white}{GACTCGGGACGTTG}CAATTGCT\$CTTGACTCGGG... \\
\textcolor{white}{\ GACTCGGTCGTTG}CAATTGCT\$CTTGACTCGGT...
\end{tabular}}
\end{center}
We extend these backward from the end of CAA for the length of the tunnel and consider the locations of the leaks, highlighted in red below:
\begin{center}
{\tt \begin{tabular}{l}
\ GACTCGGACGTTGCAA\textcolor{gray}{GTG...} \\
\ GACTCGGACGTTGCAA\textcolor{gray}{GTT...} \\
\ GACTCGGACGTTGCAA\textcolor{gray}{TCA...} \\
\ \ GACTCGG\textcolor{red}{A}GTTGCAA\textcolor{gray}{TCT...} \\
\ GACTCGGACGTTGCAA\textcolor{gray}{TGC...} \\
\ GACTCGGACGTTGCAA\textcolor{gray}{TGT...} \\
GACTC\textcolor{red}{G}GGACGTTGCAA\textcolor{gray}{TTGCT\$CTTGACTCGGG...} \\
\ GACTCGG\textcolor{red}{T}CGTTGCAA\textcolor{gray}{TTGCT\$CTTGACTCGGT...}
\end{tabular}}
\end{center}
We build a 2D grid with points where the leaks are --- in the 4th, 7th and 8th rows and the 8th, 9th and 12th columns from the right --- supporting right-above dominance counting queries:
\begin{center}
\resizebox{.37\textwidth}{!}{\tt \,G\ A\ C\ T\ C\ G\ G\ A\ C\ G\ T\ T\ G\ C\ A\ A} \\[0.5ex]
\includegraphics[width=.4\textwidth]{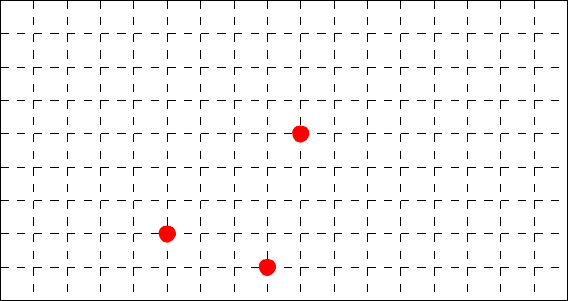}
\end{center}
We consider only the rightmost leak in each row, because once a string has left a tunnel any other differences are no longer problematic and so not true leaks.

Suppose we enter the tunnel following along the $i$th row and our pattern matches the tunnel's label until and including the label's $j$th character from the left (if our pattern doesn't match the tunnel's label at all then obviously we should not use the tunnel).  That is, suppose the $i$th occurrence of the tunnel's label in lexicographic order by the following repeated cyclic shift precedes and an occurrence of the suffix of the pattern after the tunnel's label but the $(i - 1)$st does not.  We should drop back into the BWT at the pointer for the $j$th column --- the start of the interval for the prefix of length $j$ of the tunnel's label --- minus 1 plus however many rows up to the $i$th have not yet diverged from the tunnel.  This last number is $i$ minus the number of points above and to the right of $(i, j)$ on the grid.

For example, if our pattern is {\tt GACGTTGCAATCA} then we enter the tunnel along the 3rd row and match the tunnel's label for 10 characters, to the 7th column from the left.  There are no points on the grid above and to the right of $(3, 7)$ and the BWT interval for the suffix {\tt GACGTTGCAA} of the tunnel's label is $\BWT [96..101]$, so we drop back into the BWT at $\BWT [96 - 1 + 3 = 98]$.  If our pattern is {\tt GACGTTGCAATGT...} then we enter the tunnel along the 6th row, however, and there is single point $(4, 9)$ above and to the right of $(6, 7)$ and so we drop back into the BWT at $\BWT [96 - 1 + 5 = 100]$.

This gives us our patching lemma below.  It is probably still hard to tunnel optimally with overlapping tunnels even allowing patching, and it may even be hard to tunnel with non-overlapping tunnels when we can patch tunnels, but in practice it should be useful to tunnel well-conserved regions in genomes even when there are a few variations in them.

\begin{lemma}
We can augment a BWT-based index with a patched tunnel with $\ell$ leaks at the cost of using extra space proportional to the tunnel's length plus $\ell$.  The query times for the patched tunnel are asymptotically the same as for an unpatched tunnel.
\end{lemma}

%\bibliographystyle{plain}
%\bibliography{patching}

\begin{thebibliography}{10}

\bibitem{AGNS19}
Jarno Alanko, Travis Gagie, Gonzalo Navarro, and Louisa {Seelbach Benkner}.
\newblock Tunneling on wheeler graphs.
\newblock In {\em Proc.\ Data Compression Conference (DCC)}, 2019.

\bibitem{Bai18}
Uwe Baier.
\newblock On undetected redundancy in the {Burrows-Wheeler Transform}.
\newblock In {\em Proc.\ 29th Symposium on Combinatorial Pattern Matching
  (CPM)}, 2018.

\bibitem{BD19}
Uwe Baier and Kadir Dede.
\newblock {BWT} tunnel planning is hard but manageable.
\newblock In {\em Proc.\ Data Compression Conference (DCC)}, 2019.

\bibitem{BSL26}
Nathaniel~K Brown, Vikram~S Shivakumar, and Ben Langmead.
\newblock Col-bwt: Pangenomic seed chaining with maximal matches improves read
  classification.
\newblock {\em Journal of Computational Biology}, 2026.

\bibitem{BW94}
Michael Burrows and David~J. Wheeler.
\newblock A block-sorting lossless data compression algorithm.
\newblock Technical Report 124, Digital Equipment Corportation Systems Research
  Center, 1994.

\bibitem{Dur14}
Richard Durbin.
\newblock Efficient haplotype matching and storage using the positional
  {Burrows--Wheeler transform} ({PBWT}).
\newblock {\em Bioinformatics}, 2014.

\bibitem{FM05}
Paolo Ferragina and Giovanni Manzini.
\newblock Indexing compressed text.
\newblock {\em Journal of the ACM}, 2005.

\bibitem{GMS17}
Travis Gagie, Giovanni Manzini, and Jouni Sir{\'e}n.
\newblock Wheeler graphs: A framework for {BWT}-based data structures.
\newblock {\em Theoretical Computer Science}, 2017.

\bibitem{GB22}
Adri{\'a}n Goga and Andrej Bal{\'a}{\v{z}}.
\newblock Prefix-free parsing for building large tunnelled {Wheeler} graphs.
\newblock In {\em Proc.\ 22nd Workshop on Algorithms in Bioinformatics (WABI)},
  2022.

\bibitem{MRRS07}
Sabrina Mantaci, Antonio Restivo, Giovanna Rosone, and Marinella Sciortino.
\newblock An extension of the {Burrows--Wheeler} transform.
\newblock {\em Theoretical Computer Science}, 2007.

\bibitem{ZBGL26}
Mohsen Zakeri, Nathaniel~K Brown, Travis Gagie, and Ben Langmead.
\newblock Movi 2: fast and space-efficient queries on pangenomes.
\newblock {\em Bioinformatics}, 2026.

\end{thebibliography}

\end{document}